\documentclass[lincst,sechang,a4paper]{svmultln}
\usepackage[latin1]{inputenc}
\usepackage{amsmath}
\usepackage{amsfonts}
\usepackage{amssymb}
\usepackage{graphicx}
\usepackage{makeidx}
\usepackage{adjustbox}

\begin{document}
\title{Determining Training Needs for Cloud Infrastructure Investigations using I-STRIDE}
\titlerunning{Cloud Investigation Training using I-STRIDE}

\author{}
\institute{}
\author{Joshua I. James\inst{1} \and Ahmed F. Shosha\inst{2} \and Pavel Gladyhsev\inst{2}}
\authorrunning{Joshua I. James et al.}
\toctitle{Determining Training Needs for Cloud Infrastructure Investigations using I-STRIDE}
\tocauthor{Joshua I. James, Ahmed F. Shosha, Pavel Gladyshev}
\institute{Digital Forensic Investigation Research Group: ASP Region\\
Soon Chun Hyang University\\
Shinchang-myeon, Asan-si, South Korea\\
\and
Digital Forensic Investigation Research Group: Europe\\
University College Dublin\\
Belfield, Dublin 4, IE}

\maketitle

\begin{abstract}
As more businesses and users adopt cloud computing services, security vulnerabilities will be increasingly found and exploited. There are many technological and political challenges where investigation of potentially criminal incidents in the cloud are concerned. Security experts, however, must still be able to acquire and analyze data in a methodical, rigorous and forensically sound manner. This work applies the STRIDE asset-based risk assessment method to cloud computing infrastructure for the purpose of identifying and assessing an organization's ability to respond to and investigate breaches in cloud computing environments. An extension to the STRIDE risk assessment model is proposed to help organizations quickly respond to incidents while ensuring acquisition and integrity of the largest amount of digital evidence possible. Further, the proposed model allows organizations to assess the needs and capacity of their incident responders before an incident occurs.

\end{abstract}

\section{Introduction}
New concepts in cloud computing have created new challenges for security teams and researchers alike \cite{Vouk2008}. Cloud computing service and deployment models have a number of potential benefits for businesses and customers, but security and investigation challenges -- some inherited from `traditional' computing, and some unique to cloud computing -- create uncertainty and potential for abuse as cloud technologies proliferate.

According to a survey from Ponemon Institute \cite{Ponemon2011}, only 35\% of IT respondents and 42\% of compliance respondents believe their organizations have adequate technologies to secure their Infrastructure as a Service (IaaS) environments. The report shows that respondents believe IaaS is less secure than their on-premise systems, however, ``[m]ore than half (56 percent) of IT practitioners say that security concerns will not keep their organizations from adopting cloud services''. A drive towards cloud service offerings is reiterated by Gartner \cite{Gartner2012}, who forecasts that spending on cloud computing at each service model layer will more than double by 2016. At the same time Ernst and Young \cite{EY2011} found that there is a perceived increase in risk by adopting cloud and mobile technologies, and many respondents believe that these risks are not currently being adequately dealt with. However, North Bridge \cite{Skok2012} suggests that confidence in cloud computing is increasing, even though maturity of the technologies remains a concern.

An increased confidence in cloud computing and a drive to improve business processes while reducing costs are leading to security of such systems sometimes being a secondary concern. This attitude has somewhat been carried over from traditional computing, which could possibly result in the same, or similar, security challenges, such as those presented by the Computer Research Association \cite{CRA2003} in the \textit{Four Grand Challenges in Trustworthy Computing}. If both security and insecurity from traditional computing are inherited by cloud computing, both may be augmented with the increased complexity of the cloud model, the way that services are delivered, and on-demand extreme-scale computing. Each cloud deployment and service model has its own considerations as far as security and liability are concerned. For example, in a private, single-tenant cloud where all services may be hosted on-premise, the risks are similar to on-premise, non-cloud hosting. The organization has end-to-end control, can implement and target security systems, and can control critical data flow and storage policies. A challenge with this model is that the organization must have the capability to be able to create, implement, and maintain a comprehensive security strategy for increasingly complex systems.

Several works have previously examined some cloud security concerns \cite{Armbrust2009, Jansen2011, Balduzzi2012, Kui2012, Zissis2012, Dykstra2012, Ruan2013}. This work, however, is concerned with an organization's ability to assess the investigation and response capability of their investigators considering the organization's unique needs. Security groups, such as the National Institute of Standards and Technology, have previously called for digital forensic readiness to be included in incident response planning \cite{Kent2006}. However, determining the required capabilities of investigators has not been directly addressed. Prior works, such as Kerrigan \cite{Kerrigan2013} proposed a capability maturity model for digital investigations. Such maturity models essentially focus on assessing how standardized knowledge and processes are in a particular organization, and how well these organizations actually conform to these standards. While technical capability of digital investigators is a factor, this model does not include assessment of specific technical needs of an organization. Pooe and Labuschange \cite{Pooe2012} proposed a model for assessing digital forensic readiness that includes identification of more specific technical challenges; however, this model too does not help guide an organization in specifically defining their internal digital investigation capability needs in regards to the technical skills.

\subsection{Contribution}
This work proposes a method to guide organizations in determining the technical skills needed for incident response and digital investigations that are tailored specifically to the organization. The proposed method uses an extension of a previously known asset-based risk assessment model to help guide an organization and prepare for digital investigations, including determination of technical training that is specifically required for investigators within the organization.

The remainder of this paper first discusses related prior work for assessing risk to cloud infrastructure. After, a method is described for assessing investigation capability based on an organizational risk assessment. A case study is then given applying the proposed model to cloud infrastructure. In-house knowledge can be questioned based on the identification and prioritization of risks, and gaps in knowledge may be identified from which specified training areas can be defined. Finally, conclusions are given and potential future work is discussed. 

\section{Assessing Risk to Cloud Infrastructure}
To help in the identification of threats, their impact on a system, potential evidential traces and technical skill needed by investigators, an extension to the six-category, threat categorization model -- `Spoofing, Tampering, Repudiation, Information Disclosure, Denial of Service, and Elevation of Privilege' (STRIDE) \cite{Swiderski2004} -- is proposed. The STRIDE model is a threat categorization model that can be used to help understand the impact of a specific threat being exploited in a system \cite{MSDN2005}. It helps to determine vectors of attack, the impact of an attack on data, and the overall impact to the organization due to the altered - or loss of - data. The STRIDE model has previously been applied to probabilistic risk assessment in cloud environments \cite{Saripalli2010}, threat modeling using fuzzy logic \cite{Sodiya2007}, among others.

James, Shosha et al. \cite{James2013} previously proposed an extension to the STRIDE model beyond risk assessment and potential exploitation results, to add the identification of possible investigation-relevant traces produced by the exploitation, named the ``Investigation STRIDE model'', or I-STRIDE.

\begin{figure}
\centering
\includegraphics[width=0.8\textwidth]{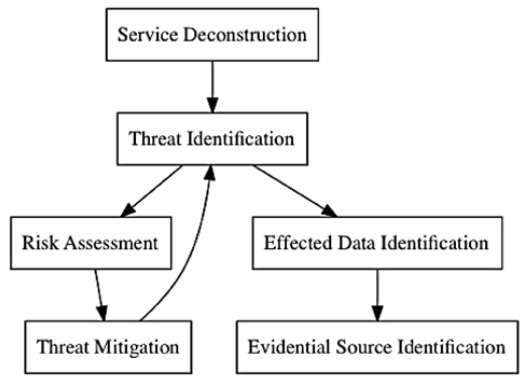}
\caption{The I-STRIDE process model for risk assessment, mitigation and investigation}
\label{fig:ISTRIDE model}
\end{figure}

As shown in Figure \ref{fig:ISTRIDE model}, the I-STRIDE process is conducted by first deconstructing a service into its dependent components. A risk assessment is conducted per component, and risk mitigation techniques are derived. Each risk identified by I-STRIDE has associated investigation-relevant data sources. When a threat to a component has been identified, an investigator may determine what data is likely to be effected by the threat. From this subset of affected data, specific data sources that may be of evidential value can be identified. These potential evidential data sources may then be used for pre-investigation planning and data targeting purposes.

\begin{figure}
\centering
\includegraphics[width=0.8\textwidth]{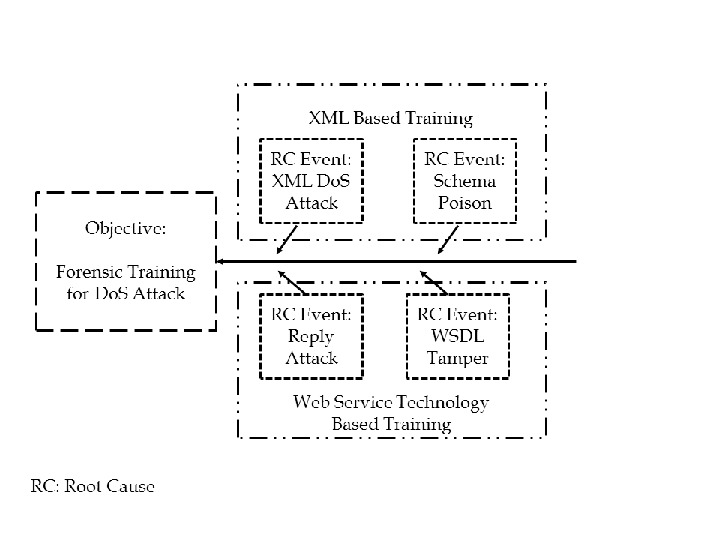}
\caption{Root Cause Analysis to guide forensic investigation training Stephenson, Peter. "Modeling of post-incident root cause analysis." International Journal of Digital Evidence 2.2 (2003): 1-16.}
\label{fig:RCA_Analysis}
\end{figure}

Determining forensic investigation knowledge required to investigate a particular threat can be modeled using risk analysis and assessment techniques. In particular, risk assessment models such as Root Cause Analysis (RCA) \cite{Stephenson2003} can be used to help identify required training. Broadly speaking, RCA is used to identify the root cause of an event that causes a phenomena of interest. Thus, RCA in combination with the proposed I-STRIDE model can be used as a basis to identify training related to the investigation of identified threats (Figure \ref{fig:RCA_Analysis}).

Utilizing methodologies such as RCA benefits not only the process of training identification, or gaps in knowledge, but also training efficacy.
When considering investment in training, organizations attempt to determine whether a specific training meets their unique needs. By identifying gaps in knowledge related to prioritized organizational risks, training can be more focused at areas the organization specifically needs. 
As such, training can be audited according to the identified training objectives and scope based on the needs of the organization.

\section{Incident Response Planning and Capability Assessment}
An important step in Incident Response -- if not the most important -- is the readiness phase. In the integrated digital investigation process (IDIP) model, Carrier and Spafford \cite{Carrier2003} state that ``the goal of the readiness phases is to ensure that the operations and infrastructure are able to fully support an investigation''. The operations readiness phase involves the on-going training of personnel, such as first responders and lab technicians, and the procurement and testing of equipment needed for the investigation. However, while general training may be applicable to each organization, an organization may have specific training needs that need to be identified.

For example, operational readiness in cloud environments should include education in cloud-related technologies, such as hypervisors, virtual machines and cloud-based storage, but may specifically depend on what services the organization is providing. Personnel should have general knowledge of how to interact with cloud technologies at the infrastructure, platform and software layers, and understand the effect their actions have on the environment. They should understand the methods and tools available to collect investigation-relevant data in each layer of the cloud. Different Cloud Service Providers (CSP) may have proprietary systems, so training on the use and investigation of these proprietary systems should be considered. However, determination of exactly what skills and knowledge are necessary to ensure quality investigations may be difficult. Further, identifying what technologies should be the focus of technical training may not be fully known.

The training of personnel, identification of potential risks, and identification of potential data sources before an incident occurs can greatly help in efficient incident response, and with the timely and sound acquisition of relevant data. For this reason this work recommends organizations model threats, their potential impact, and potential evidential trace data sources before an incident occurs. This will assist the CSP in preserving potential evidence during incident response, and will help law enforcement have a better idea of what data will be available, and how to handle such data, if a particular incident occurs.

\section{Methodology}
\label{sec:Methodology}
The proposed knowledge identification method is broken into two areas of assessment: Technology (security) risk and Knowledge (training/education) risk. Assessment of `knowledge risk' is necessary because simply knowing a technical vulnerability exists will not aid in incident response or investigation unless the responder/investigator has knowledge of concepts such as where relevant evidential data may exist and how to properly acquire such data. Below are both the Technology risk and knowledge risk assessment processes.

\begin{itemize}
\item Technology Risk Assessment (I-STRIDE)
\begin{enumerate}
\item Identify Assets
\item Identify Threats to Assets
\item Determine Potential Threat Impact
\item Determine Potential Evidential Data Sources (Pre-Investigation)
\item Prioritize Threats
\begin{itemize}
\item Organizational needs
\item Common Vulnerability Scoring System (CVSS) \cite{NIST}
\end{itemize}
\end{enumerate}
\item Knowledge Risk Assessment
\begin{enumerate}
\item Identify required investigation knowledge
\item Assess current in-house knowledge
\begin{itemize}
\item Knowledge of the collection/analysis of associated evidential data sources
\end{itemize}
\item Compare in-house knowledge with risk prioritization
\end{enumerate}
\end{itemize}

Knowledge risk in this case can be assessed based on Bloom's Taxonomy \cite{Anderson2005a}. Using Bloom's Taxonomy in-house knowledge could be assessed, either through self-assessment or a more formal process. The Taxonomy would allow an organization to understand the level of knowledge they possess about the investigation of breaches caused by the specific vulnerability.

Bloom's Taxonomy has 6 `verbs' that correspond to a level of knowledge: remembering, understanding, applying, analyzing, evaluating, and creating. Knowledge of a vulnerability or evidential data sources can be assessed based on these levels of knowledge. As such, a score can be assigned to each level (1 - 6), which can be considered the knowledge risk score. Such a score implies that higher-level knowledge such as an ability to analyze and evaluate a topic is preferred over simple remembering.

If an organization is also using a threat prioritization metric -- such as the CVSS -- then these scores can be combined to create a `knowledge prioritization' model. For example, with CVSS a score of 1 indicates a low risk, where a score of 10 indicates a high risk. In Bloom's taxonomy a score of 1 indicates low knowledge, and a score of 6 indicates higher-level knowledge. Assume CVSS is used to assess the severity of a threat (1 being least severe, 10 being the most severe). Bloom's taxonomy measures can be scaled to the same scale as CVSS, and the knowledge risk can be subtracted from the technology risk as so:

$ T_s - ((K_s\div K_{max})\cdot T_{max}) $

\noindent where:
\begin{itemize}
\item $T_s$ is the technology risk score
\item $K_s$ is the knowledge risk score
\item $K_{max}$ is the maximum knowledge risk score
\item $T_{max}$ is the maximum technology risk score
\end{itemize}

In this case, if the technology risk score is high (10 out of 10), and the knowledge risk score is also high (6 out of 6), then the \textit{knowledge priority} will be low (0) in terms of training or education needs. If the technology risk score is low (2 out of 10), and the knowledge risk score is also low (1 out of 6), then the overall knowledge priority will still remain low (0.33). The threats with the highest priority will be high-scoring technology risks that the organization has little knowledge about. Further, as knowledge is updated (either gained or lost) knowledge risk can also be updated to reflect the current state of knowledge in the organization.

Again, this prioritization is used to identify areas where investigation education or training is lacking and supplementation may be necessary due to technology risk, not to imply that a high knowledge of a vulnerability will reduce an organization's risk of that vulnerability being exploited.

\section{Case Studies}
To show the applicability of the I-STRIDE model for determining training needs, assessment of cloud computing infrastructure based on Eucalyptus \cite{Eucalyptus2013} and OpenStack will be given as examples.

\subsection{Case 1: Eucalyptus Cloud}
This case will specifically look at a deployed Eucalyptus Cloud. The components of this platform will be explained, and an analysis using the I-STRIDE model will be conducted against this deployment.

The Eucalyptus architecture is composed of five high-level components that are essentially standalone web services. These components include:

\begin{itemize}
\item Cloud Controller (CLC): The cloud controller is the main entry point for the cloud environment. CLC is responsible for ``exposing and managing the underlying virtualized resources''.
\item Cluster Component (CC): CC is responsible for managing the execution of VM instances.
\item Storage Controller (SC): Provides block-level network storage that can be dynamically attached by VMs instances.
\item Node Controller (NC): Executed on every node that is designated for hosting and allows management of VM instances.
\item Walrus: Allows the storage and management of persistent data.
\end{itemize}

\begin{figure}
\centering
\includegraphics[width=0.8\textwidth]{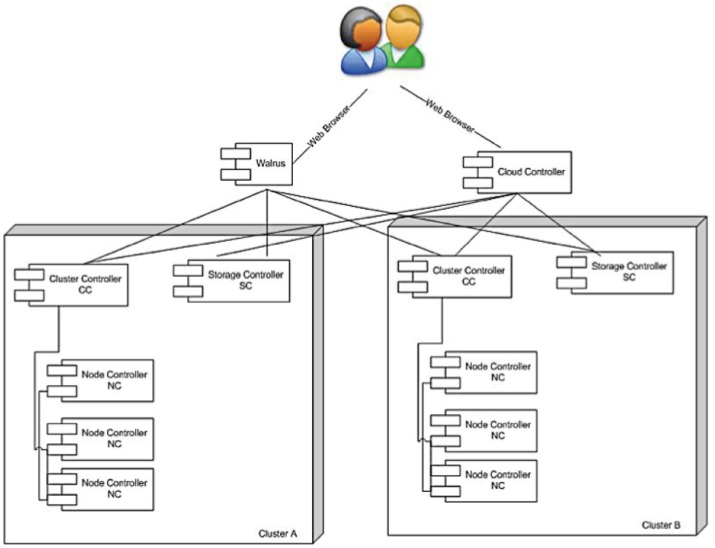}
\caption{Deployed Eucalyptus Architecture}
\label{fig:deployed_eucalyptus}
\end{figure}

Figure \ref{fig:deployed_eucalyptus} shows the Eucalyptus components and their connection and communication channels.

The scope of this case will be limited to asset-centric threat modeling. The assets in this case will be defined as each of the Eucalyptus components which can be thought of as the Cloud Service Provider (CSP), and will also include a cloud client. In this case, threats (Table \ref{tab:ISTRIDE}) were identified and exploited in the deployed Eucalyptus architecture. Per the I-STRIDE model, an analysis of affected assets was conducted. An investigation was then conducted to determine potential evidential data sources. Identified threats, the threat description, the affected asset, the impact on the asset, and the location of potential evidential data sources are listed in Table \ref{tab:ISTRIDE}. Information in this table may normally be used to assess threats, and, if a threat is exploited, help to collect potential evidential traces during the incident response phase.

\begin{table}
\centering
\begin{adjustbox}{width=\textwidth,totalheight=\textheight,keepaspectratio}
\begin{tabular}{|p{5em}|p{11em}|p{7em}|p{6em}|p{9em}|}
\hline
\textbf{Threat}&
\textbf{Description}&
\textbf{Asset}&
\textbf{Threat Impact}&
\textbf{Potential Evidential Sources}\\
\hline
XML Denial of Service&
Attacker crafts XML message with a large payload, recursive content or with malicious DTD schema.&
Cloud Controller, Cloud Client&
Denial of Service&
XML parser logs at the cloud controller\\
\hline
Replay Attack Flaws&
Attacker could issue recurrence overloaded Simple Object Access Protocol (SOAP) messages over HTTP to overwhelm the CSP and stop the service&
Cloud Controller, Cloud Client&
Denial of Service&
SOAP message timestamp and message payload to identify the message flaws\\
\hline
WSDL Parameter Tampering&
Attacker could embed command line code into WSDL documents or command shell to execute the command&
Cloud Controller, Cluster Controller, Node Controller, Cloud client&
Denial of Service&
Detailed investigation of WSDL file could identify parameter tampering\\
\hline
Schema Poisoning&
Attacker could compromise the XML schema grammar and manipulate the data&
Cloud Controller, Cloud Client&
Denial of Service&
Detailed investigation of XML parser logs at the cloud controller may contain evidence of XML schema tampering\\
\hline
\end{tabular}
\end{adjustbox}
\caption{Identified threats, the estimated threat impact and potential evidential data sources identified for a Eucalyptus that is the result of the proposed I-STRIDE model}
\label{tab:ISTRIDE}
\end{table}

Using the I-STRIDE model could help CSPs and law enforcement identify an investigation starting point during Incident Response (IR). This level of readiness would potentially allow for improved pre-planning, first response and cooperation once an incident occurred. While I-STRIDE may help to determine data sources interesting to an investigation, this information is only useful if the responders understand how to properly access, acquire and preserve such data. The output of the I-STRIDE model can be considered the knowledge the organization needs. For example, if the Cluster Component is compromised then incident responders need knowledge of the Cluster Component to be able to make use of the knowledge of associated evidential data sources. The I-STRIDE model can be used to guide training and education development plans based on the needs of an assessed organization.

\subsubsection{Case 1: Knowledge Prioritization}
Consider again Table \ref{tab:ISTRIDE}. If denial of service was found to be a priority risk in an organization, then at least 4 threats have an impact defined as denial of service. The threat vector, and already known potential evidential sources can be used to define the type of training necessary to thoroughly investigate breaches using such vectors of attack. For example, the first priority threat as shown in Table \ref{tab:invest_training} uses XML as an attack vector. XML parser logs were identified as potential evidential sources at the cloud controller.

The organization can apply the methodology described in Section \ref{sec:Methodology} to determine knowledge risk. Notice, in table \ref{tab:invest_training} the organization did not use CVSS, but instead chose a low-medium-high technology risk prioritization scheme. In this case, since each technology risk is a high priority (3 out of 3), the organization can now assess their knowledge about the investigation of each technology risk. For this case, let's assume that the organization has a great understanding of XML attack vector investigations (5 out of 6), and very little knowledge of SOAP exploit prevention and investigation (1 out of 6). Knowledge prioritization can then be assessed as follows
\begin{itemize}
\item XML investigation training priority: $3 - ((5\div 6)\cdot 3) = 0.5$
\item SOAP investigation training priority: $3 - ((1\div 6)\cdot 3) = 2.5$
\end{itemize}

If an organization can answer questions about in-house knowledge for high priority technology risks, then the organization can identify associated knowledge risk, and invest in training of personnel more effectively based on their unique needs. Once these knowledge priority areas have be en identified, they can be fed directly into training development models such as the Successive Approximation Model \cite{Allen2006} for rapid training development. This will allow organizations to quickly target and close gaps in knowledge based on prioritized organizational risks.

\begin{table}
\centering
\begin{adjustbox}{width=\textwidth,totalheight=\textheight,keepaspectratio}
\begin{tabular}{|c|p{5em}|p{9em}|p{5em}|p{9em}|p{6em}|}
\hline
\textbf{Priority}&
\textbf{Threat}&
\textbf{Asset}&
\textbf{Threat Impact}&
\textbf{Potential Evidential Sources}&
\textbf{Knowledge}\\
\hline
High&
XML Denial of Service&
Cloud Controller, Cloud Client&
Denial of Service&
XML parser logs at the cloud controller&
XML attack vector investigation\\
\hline
High&
Replay Attack Flaws&
Cloud Controller, Cloud Client&
Denial of Service&
SOAP message timestamp and message payload to identify the message flaws&
SOAP exploit prevention and investigation\\
\hline
High&
WSDL Parameter Tampering&
Cloud Controller, Cluster Controller, Node Controller, Cloud client&
Denial of Service&
Detailed investigation of WSDL file could identify parameter tampering&
WSDL Security and Investigation\\
\hline
High&
Schema Poisoning&
Cloud Controller, Cloud Client&
Denial of Service&
Detailed investigation of XML parser logs at the cloud controller may contain evidence of XML schema tampering&
XML attack vector investigation\\
\hline
\end{tabular}
\end{adjustbox}
\caption{Threat prioritization (in this case, based on the organization's subjective decision) and required investigation knowledge identification based on identified threats that cause a particular class of threat impact}
\label{tab:invest_training}
\end{table}

\subsection{Case 2: OpenStack}
The next case concerns the assessment of OpenStack, an open source cloud infrastructure project. This example will use vulnerabilities identified in CVE Details \cite{MITRE_CVE}, along with the threat's identified CVSS score. From CVE, an organization running OpenStack may assess their specific technology and knowledge risk in relation to new vulnerabilities.

\begin{table}
\centering
\begin{adjustbox}{width=\textwidth,totalheight=\textheight,keepaspectratio}
\begin{tabular}{|p{5em}|p{11em}|p{7em}|p{5em}|p{4em}|p{7em}|p{6em}|}
\hline
\textbf{Threat}&
\textbf{Description}&
\textbf{Asset}&
\textbf{Threat Impact}&
\textbf{CVSS}&
\textbf{Potential Evidential Sources}&
\textbf{Knowledge}\\
\hline
Issue requests with an old X-Timestamp value&
Authenticated attacker can fill an object server with superfluous object tombstones&
Swift Cluster&
Denial of Service&
4.0&
Tombstone files&
Swift Object Servers\\
\hline
Re-auth deleted user with old token &
When an actor claims to have a given identity, the software does not prove or insufficiently proves that the claim is correct.&
Keystone&
Security Bypass&
6.0&
Instance and user logs&
Swift Proxy\\
\hline
Generate unparsable or arbitrary XML responses&
Unchecked user input in Swift XML responses&
Swift account servers&
Security Bypass&
7.5&
Account server logs&
Account server\\
\hline
\end{tabular}
\end{adjustbox}
\caption{Identified threats, the estimated threat impact and potential evidential data sources identified for OpenStack that is the result of CVE details and the I-STRIDE model}
\label{tab:OpenStack_ISTRIDE}
\end{table}

The knowledge required for each of the technology risks identified in Table \ref{tab:OpenStack_ISTRIDE} can be identified using the I-STRIDE process, and specifically by simulating an incident and determine where data relevant to the investigation may be found. In this case, the required knowledge has been defined as parts of the swift architecture.

The CVSS in this case represents the technology risk if the vulnerability has not been patched yet, or for some reason cannot be. The organization must conduct an internal knowledge risk assessment based on the identified knowledge areas for newly identified vulnerabilities. In this case, assume an organization has moderate (3 out of 6) knowledge about swift proxy and account servers, and little (1 out of 6) knowledge of swift object servers. The investigation knowledge priority for each threat can be calculated as so:

\begin{itemize}
\item Object server investigation training priority: $4 - ((1\div 6)\cdot 10) = 2.33$
\item Proxy server investigation training priority: $6 - ((3\div 6)\cdot 10) = 1$
\item Account server investigation training priority: $7.5 - ((3\div 6)\cdot 10) = 2.5$
\end{itemize}

In this case, because the account server has the highest technology risk and the organization only has a moderate level of knowledge about the investigation of such a risk, it is given the highest priority. It is then followed by a technology risk that is relatively low, but is a risk which the organization does not have much knowledge about.

\section{Conclusions}
Cloud computing has a number of benefits, such as high availability, potentially lower cost, and potentially improved security. However, cloud computing also has a number of associated risks. Some of these risks have been inherited from traditional computing models, while the cloud business model introduces others. As more businesses and end users move their data and processing to cloud environments, these environments will increasingly become the target, or even the originator, of malicious attacks. By taking an asset-based risk assessment approach, and specifically using the I-STRIDE model, organizations can identify and prioritize threats, determine threat impact and potential evidential sources, and ultimately identify gaps in investigator knowledge before a threat is exploited. By implementing the proposed technology and knowledge risk assessment metrics, and organization can at least be better positioned to make training and education investment decisions based on observed deficiencies. I-STRIDE can act as a base for CSPs and law enforcement to more effectively work together before and during the investigation of incidents in cloud environments, and not only in the discussion of vulnerabilities but in the discussion of required knowledge.

While this work proposed a naive model for determining investigator training needs specific to the organization, future work will attempt to evaluate the model with real organizations rather than a researcher-created case study. For example, the model, as proposed, integrates prevention (security) and investigation (post incident) to attempt to improve both. However, such a model takes a considerable amount of effort and pre-planning to implement. In large organizations, even communication between investigators and security officers may be difficult. Such real-world case studies are needed to evaluate the practicality of the proposed training-guidance method.

\bibliographystyle{plain}
\bibliography{ISTRIDE}

\begin{thebibliography}{10}

\bibitem{Allen2006}
Michael~W. Allen.
\newblock {\em {Creating successful e-learning: a rapid system for getting it
  right the first time, every time}}.
\newblock Pfeiffer \& Co., 2006.

\bibitem{Anderson2005a}
Lorin~W. Anderson, David~R. Krathwohl, and Benjamin~Samuel Bloom.
\newblock {\em {A taxonomy for learning, teaching, and assessing}}.
\newblock Longman, 2005.

\bibitem{Armbrust2009}
Michael Armbrust, Armando Fox, Rean Griffith, Anthony~D Joseph, Randy~H Katz,
  Andrew Konwinski, Gunho Lee, David~A Patterson, Ariel Rabkin, Ion Stoica, and
  Matei Zaharia.
\newblock {Above the Clouds: A Berkeley View of Cloud Computing}.
\newblock {\em Science}, 53(UCB/EECS-2009-28):07--013, 2009.

\bibitem{Balduzzi2012}
Marco Balduzzi, Jonas Zaddach, Davide Balzarotti, Engin Kirda, and Sergio
  Loureiro.
\newblock {A security analysis of amazon's elastic compute cloud service},
  2012.

\bibitem{Carrier2003}
Brian~D Carrier and Eugene~H Spafford.
\newblock {Getting physical with the digital investigation process}.
\newblock {\em International Journal of Digital Evidence}, 2(2):1--20, 2003.

\bibitem{CRA2003}
CRA and Computing~Research Association.
\newblock {Four Grand Challenges in Trustworthy Computing}.
\newblock Technical report, 2003.

\bibitem{Dykstra2012}
Josiah Dykstra and Alan~T Sherman.
\newblock {Acquiring forensic evidence from infrastructure-as-a-service cloud
  computing: Exploring and evaluating tools, trust, and techniques}.
\newblock {\em Digital Investigation}, 9:S90--S98, August 2012.

\bibitem{Eucalyptus2013}
Eucalyptus.
\newblock {Eucalyptus: The Open Source Cloud Platform}, 2013.

\bibitem{EY2011}
EY and EYGM Limited.
\newblock {Into the cloud, out of the fog: Ernst \& Young's 2011 Global
  Information Security Survey}.
\newblock Technical report, 2011.

\bibitem{Gartner2012}
Gartner.
\newblock {Forecast: Public Cloud Services, Worldwide, 2010-2016, 2Q12 Update}.
\newblock Technical report, 2012.

\bibitem{James2013}
Joshua~I James, Ahmed~F Shosha, and Pavel Gladyshev.
\newblock {Digital Forensic Investigation and Cloud Computing}.
\newblock In Keyun Ruan, editor, {\em Cybercrime and Cloud Forensics:
  Applications for Investigation Processes}, pages 1--41. IGI Global, 2013.

\bibitem{Jansen2011}
W~A Jansen.
\newblock {Cloud Hooks: Security and Privacy Issues in Cloud Computing}.
\newblock pages 1--10. IEEE, 2011.

\bibitem{Kent2006}
Karen Kent, Suzanne Chaevalier, Tim Grance, and Hung Dang.
\newblock {Guide to Integrating Forensic Techniques into Incident Response}.
\newblock Technical Report SP800-86, 2006.

\bibitem{Kerrigan2013}
Martin Kerrigan.
\newblock {A capability maturity model for digital investigations}.
\newblock {\em Digital Investigation}, 10(1):19--33, March 2013.

\bibitem{Kui2012}
Ren Kui, Wang Cong, and Wang Qian.
\newblock {Security Challenges for the Public Cloud}.
\newblock {\em Internet Computing, IEEE}, 16(1):69--73, 2012.

\bibitem{MITRE_CVE}
MITRE.
\newblock {OpenStack Security Vulnerabilities}.

\bibitem{MSDN2005}
MSDN.
\newblock {The STRIDE Threat Model}, 2005.

\bibitem{NIST}
NIST.
\newblock {Common Vulnerability Scoring System}.

\bibitem{Ponemon2011}
Ponemon and L~L~C {Ponemon Institute}.
\newblock {The Security of Cloud Infrastructure: Survey of U.S. IT and
  Compliance Practitioners}.
\newblock Technical report, 2011.

\bibitem{Pooe2012}
Antonio Pooe and L~Labuschagne.
\newblock {A conceptual model for digital forensic readiness}.
\newblock In {\em 2012 Information Security for South Africa}, pages 1--8.
  IEEE, August 2012.

\bibitem{Ruan2013}
Keyun Ruan, Joe Carthy, Tahar Kechadi, and Ibrahim Baggili.
\newblock {Cloud forensics definitions and critical criteria for cloud forensic
  capability: An overview of survey results}.
\newblock {\em Digital Investigation}, 10(1):34--43, June 2013.

\bibitem{Saripalli2010}
P~Saripalli and B~Walters.
\newblock {QUIRC: A Quantitative Impact and Risk Assessment Framework for Cloud
  Security}.
\newblock pages 280--288. Ieee, 2010.

\bibitem{Skok2012}
Michael~J Skok.
\newblock {Future of Cloud Computing 2012}, 2012.

\bibitem{Sodiya2007}
Adesina~Simon Sodiya, Saidat~Adebukola Onashoga, and Beatrice Oladunjoye.
\newblock {Threat modeling using fuzzy logic paradigm}.
\newblock {\em Journal of Issues in Informining Science and Technology},
  4(1):53--61, 2007.

\bibitem{Stephenson2003}
Peter Stephenson.
\newblock {Modeling of post-incident root cause analysis}.
\newblock {\em International Journal of Digital Evidence}, 2(2):1--16, 2003.

\bibitem{Swiderski2004}
F~Swiderski and W~Snyder.
\newblock {\em {Threat modeling}}.
\newblock Microsoft Press Redmond, WA, USA, 2004.

\bibitem{Vouk2008}
M~A Vouk.
\newblock {Cloud computing-Issues, research and implementations}.
\newblock pages 31--40. IEEE, 2008.

\bibitem{Zissis2012}
Dimitrios Zissis and Dimitrios Lekkas.
\newblock {Addressing cloud computing security issues}.
\newblock {\em Future Generation Computer Systems}, 28(3):583--592, 2012.

\end{thebibliography}

\end{document}